\documentclass[11pt,eqsecnum,floats,nofootinbib,amsmath,amssymb, prd]{revtex4}
\usepackage{setspace}
\usepackage{slashed}

\usepackage{tensor}\newcommand{\ou}[3]{{#1}{}^{#2}{}_{#3}}
\newcommand{\uo}[3]{{#1}{}_{#2}{}^{#3}}
\newcommand{\prefix}[2]{\tensor[^{\mathnormal{#1}}]{#2}{}}
\newcommand{\po}[3]{\tensor[^{\mathnormal{#1}}]{#2}{^{#3}}}
\newcommand{\pu}[3]{\tensor[^{\mathnormal{#1}}]{#2}{_{#3}}}
\newcommand{\pou}[4]{\tensor[^{\mathnormal{#1}}]{#2}{^{#3}_{#4}}}
\newcommand{\puo}[4]{\tensor[^{\mathnormal{#1}}]{#2}{_{#3}^{#4}}}
\newcommand{\q}[1]{`#1'\,}  

\usepackage[normalem]{ulem}

\usepackage{amsfonts,amsthm,amscd, bbm}
\usepackage{graphicx}
\usepackage{enumerate} 
\usepackage{colordvi} 
\usepackage{units}
\usepackage{epsfig}
\usepackage{color}
\usepackage[english]{babel}
\usepackage[bookmarksopen]{hyperref}
\usepackage{tensor}
\usepackage{leftidx}
\usepackage[T1]{fontenc}
\usepackage{natbib}
\usepackage{mathtools}

\newcommand{\be}{\begin{equation}}
\newcommand{\ee}{\end{equation}}
\newcommand{\beq}{\begin{equation}}
\newcommand{\eeq}{\end{equation}}
\newcommand{\bes}{\begin{eqnarray}}
\newcommand{\ees}{\end{eqnarray}}
\newcommand{\bqa}{\begin{eqnarray}}
\newcommand{\eqa}{\end{eqnarray}}
\newcommand{\bea}{\begin{eqnarray}}
\newcommand{\eea}{\end{eqnarray}}

\newcommand{\su}{\mathfrak{su}}

\def\demi{{\frac{1}{2}}}

\newcommand{\f}{\frac}

\def\ra{{\rangle}}
\def\la{{\langle}}

\def\dr{{\rightarrow}}

\def\demi{{\frac{1}{2}}}

\DeclareMathOperator{\SU}{SU}

\def\bT{\prefix{\bullet}{T}}
\def\ciT{\prefix{\circ}{T}}
\def\bD{\prefix{\bullet}{D}}
\def\ciD{\prefix{\circ}{D}}
\def\ciR{\prefix{\circ}{R}}

\def\oo{\prefix{\bullet}{\omega}}
\def\oLC{\prefix{\circ}{\omega}}

\usepackage{hyperref}
\newcommand{\rd}{\mathrm{d}}

\newcommand{\di}{\mathrm{d}}
\newcommand{\eref}[1]{(\ref{#1})}

\begin{document}\singlespacing
\title{\Large First-order formulation of teleparallel gravity and dual loop gravity}
\author{Ma\"it\'e Dupuis}
\email{mdupuis@perimeterinstitute.ca}
\affiliation{Perimeter Institute for Theoretical Physics, Waterloo, Ontario,
Canada}
\affiliation{Department of Applied Mathematics, University of Waterloo, Waterloo, Ontario, Canada}

\author{Florian Girelli}
\email{fgirelli@uwaterloo.ca}
\affiliation{Department of Applied Mathematics, University of Waterloo, Waterloo, Ontario, Canada}

\author{Abdulmajid Osumanu}
\email{a3osumanu@uwaterloo.ca}
\affiliation{Department of Applied Mathematics, University of Waterloo, Waterloo, Ontario, Canada}

\author{Wolfgang Wieland}
\email{wwieland@perimeterinstitute.ca}
\affiliation{Perimeter Institute for Theoretical Physics, Waterloo, Ontario,
Canada}

\date{\today}

\begin{abstract}\noindent
There are at least two ways to encode gravity into geometry: Einstein's general theory of relativity (GR) for the metric tensor, and  teleparallel gravity, where torsion as opposed to curvature encodes the dynamics of the gravitational degrees of freedom.  The main purpose of the paper is to explore the relation between loop gravity and teleparallel gravity. 
We argue that these two formulations of gravity  are related to two different discretizations of the Einstein--Cartan action, which were studied recently in the literature. The first discretization leads to the \emph{{loop gravity}} kinematical phase space where the zero torsion condition is enforced {first} and the other is the \emph{{dual loop gravity}} kinematical phase space where  curvature is imposed to vanish {first}. Our argument is based on the observation that the GR first-order Einstein--Cartan action can also be seen as a first-order action for teleparallel gravity up to a boundary term. The results of our paper suggest that the \emph{dual loop gravity} framework  is a natural discretization of  teleparallel gravity, whereas \emph{loop gravity}  is  naturally related to the standard GR metric description.
\end{abstract}


\maketitle


\section{Introduction}
\noindent When solving the field equations for gravity, typically one deals with a first-order or a second-order formulation. In the first-order formulation, the equations of motion are first-order (partial) differential equations, whereas in the second-order formalism, the equations of motion are second-order (partial) differential equations. The two formulations are related, starting from the first-order formalism, one can eliminate some of the first-order equations of motion to obtain the second-order formalism. As a consequence, there are fewer independent variables in the second-order formulation than there are in the first-order formulation. 
 Classically, using the first-order or the second-order formalism does not really matter, since the space of initial data is the same in either formulation. At the quantum level the difference matters. Indeed, from the perspective of canonical quantization, the first-order formulation is the preferred choice, because the evolution and constraint equations are now all polynomial, and the quantization ambiguities reduce to relatively mild ordering ambiguities. In the second-order formulation, on the other hand,  the field equations are non-polynomial. They involve square roots and require the metric to be invertible,\footnote{In fact, the first-order formulation is slightly more general than the second-order formulation, because the field equations can be integrated even when the metric is degenerate.}  which adds further difficulties to the problem. These observations were  made first by Ashtekar, who showed how the Einstein--Cartan formulation leads to a more promising approach to quantum gravity than the original Wheeler--De Witt approach. Therefore, the first-order Einstein--Cartan action or some variant of it (e.g.\ the self-dual Plebanski action, the constrained BF action, or the MacDowell--Mansouri action) is usually taken as the starting point for the Hamiltonian analysis, and the loop quantum gravity (LQG) formulation  \cite{Rovelli:2004tv, Thiemann:2007pyv}.

\smallskip


There are several possibilities to encode gravity into geometry.\footnote{In the following, we will restrict ourselves to the context of \textit{metric compatible theories}. One could indeed relax the metric compatibility condition and explore how gravity can be encoded in this context. We refer to \cite{Nester:1998mp,Heisenberg:2018vsk} for more details on such generalised gravitational theories.} The first and most natural way to do so is to consider the curved metric $g_{ab}$ and the metric compatible and torsionless Christoffel connection with covariant derivative $\prefix{\circ}\nabla_a$ {: $\prefix{\circ}\nabla_a g_{ab}=0$}. For a vanishing cosmological constant  $\Lambda$, the dynamics can be derived from the familiar Einstein--Hilbert action
\be \label{actionGR}
S_{\mathrm{EH}}[g]= \frac{1}{16\pi G}\int_{\mathcal{M}} d^4v_g \ciR[g], 
\ee
where $\ciR$ is the Ricci scalar built for the metric tensor $g_{ab}$, and $d^4v_g=\sqrt{-\det g_{\mu\nu}}\,\di x^0\wedge\dots \di x^3$ is the canonical volume element. \emph{The metric formulation  is a second-order formulation} since the connection is expressed in terms of the metric and the field equations are \emph{second-order} partial differential equations.

\smallskip

By studying how half-integer spin degrees of freedom interact with gravity, Weyl introduced the concept of frame field $e$ in general relativity \cite{Weyl1929}. This led ultimately to the Sciama--Kibble--Einstein--Cartan  (SKEC) formalism for gravity. This formalism puts on equal footing both the massive and spinning degrees of freedom. For the Lorentzian $D$-dimensional case,
\be \label{actionEC}
S_{\mathrm{EC}}[e,A]= \f{1}{8\pi G} \int \big\langle B[e]\wedge F[A]\big\rangle_{\mathfrak{so}(1,D-1)},
\ee
where $F=\rd A + \frac12 [A,A]$ is the curvature of the spin connection, $B$ denotes the $(D-2)$-form $B=\ast(\wedge^{D-2}e)$, which is built from the internal hodge dual of $(D-2)$ frame fields $e^{I_1}\wedge\dots e^{I_{D-2}}$, and $\langle \cdot, \cdot \rangle_{\mathfrak{so}(1,D-1)}$ is the Killing form for $\mathfrak{so}(1,D-1)$. Our conventions are $\langle X,  Y\rangle_{\mathfrak{so}(1,D-1)}=-\f12 X^I{}_JY^J{}_I$.
On the space of histories, the $SO(1,D-1)$ connection $A$ has both non-trivial curvature and torsion.
%
\textit{The SKEC formalism is a first-order formalism for gravity}, because the equations of motion only contain first derivatives of the fundamental configuration variables. 

\smallskip

If there are no spin degrees of freedom, we get as an equation of motion that the spin connection $A$ should be torsionless. Plugging this back into the action \eqref{actionEC}, we get the Palatini action for gravity, in terms of frame fields and a torsionless spin connection. \emph{The Palatini formalism for gravity is a  second-order formalism since the equations of motion are now of second order in the metric or frame field.}

 \medskip 
 
Another way to encode gravity into geometry  \cite{einstein} is based on an affine and metric compatible flat derivative $\prefix{\bullet}{\nabla}_{a}:\prefix{\bullet}{\nabla}_{a}g_{bc}=0$. The underlying connection $\pou{\bullet}{\omega}{I}{J}$ is the so-called Weitzenb\"ock connection and it has  no curvature,
\begin{equation}
\pou{\bullet}{R}{a}{bcd}Z^bX^cY^d=\prefix{\bullet}{\nabla}_{X}\prefix{\bullet}{\nabla}_{Y}Z^a-\prefix{\bullet}{\nabla}_{Y}\prefix{\bullet}{\nabla}_{X}Z^a-\prefix{\bullet}{\nabla}_{[X,Y]}Z^a=0,
\end{equation}
where $[X,Y]^a\in TM$ is the Lie bracket between vector fields $X^a,Y^a\in TM$.
 The gravity dynamics is encoded into the associated torsion two-form $\pou{\bullet}{T}{a}{bc}$,
\begin{equation}
\pou{\bullet}{T}{a}{bc}X^bY^c=\prefix{\bullet}{\nabla}_{X}Y^a-\prefix{\bullet}{\nabla}_{Y}X^a-[X,Y]^a.
\end{equation}
The key idea behind teleparallelism is to treat the torsion two-form as the field strength of an abelian connection, which is the frame field $\ou{e}{I}{a}$,
\begin{equation}
\pou{\bullet}{T}{c}{ab}=2\uo{e}{I}{c}\,\prefix{\bullet}{\nabla}{}_{[a}\ou{e}{I}{b]},\label{Tdef}
\end{equation}
where $\prefix{\bullet}{\nabla}$ is the covariant exterior derivative with respect to the Weitzenb\"ock connection, $\prefix{\bullet}{\nabla}_aV^I=\partial_a V^I+\pou{\bullet}{\omega}{I}{Ja}V^J$. This is the teleparallel formulation, which has been slightly less explored than the GR metric formulation. For an extensive review of the theory see \cite{tele} and the references therein. The standard teleparallel action we will consider is, 
\be  \label{action1}
S_{\mathrm{TP}}[e;\prefix{\bullet}{\omega}]=  - \frac{1}{16\pi G} \int_{\mathcal{M}} d^4v_g \left(\frac{1}{4}\pu{\bullet}{T}{cab}\po{\bullet}{T}{cab}-\frac{1}{2}\prefix{\bullet}{T}_{cab}\prefix{\bullet}{T}^{abc}-\pou{\bullet}{T}{b}{ab}\pou{\bullet}{T}{ca}{c} \right).
\ee 
The field equations are satisfied when the action is stationary with respect to variations of the frame field $\ou{e}{I}{a}$.  The flat reference connection $\prefix{\bullet}{\omega}$ can be taken as an independent variable \cite{Golovnev:2017dox}, but the resulting field equations are redundant. The action defines a second-order theory for gravity, because the equations of motion derived from the action \eref{action1} are second-order partial differential equations for the frame field $\ou{e}{I}{a}$. The field equations will involve the inverse  frame field $\uo{e}{I}{a}:\uo{e}{I}{a}\ou{e}{I}{b}=\delta^a_b$, which enters the action through the definition of the torsion two-form \eref{Tdef}. From the perspective of the program of canonical quantization,  we would prefer, however, a formulation where the {(i)} field equations are polynomial and {(ii)} do not require the  co-frame fields to be invertible.\footnote{The condition $\det \ou{e}{I}{a} >0$ defines an anholonomic constraint on phase space, which is difficult to impose at the quantum level.}

\smallskip

It is well-known that the two actions, the Einstein--Hilbert action \eqref{actionGR} and the teleparallel action \eqref{action1}, are equal up to a boundary term \cite{tele},
\be \label{tel-GR}
S_{\mathrm{GR}}[g]= S_{\mathrm{TP}}[e;\prefix{\bullet}{\omega}] - \frac{1}{8\pi G}\int_{\mathcal{M}}d^4v_g\prefix{\circ}{\nabla}_a\pou{\bullet}{T}{ba}{b}.
\ee
The two actions coincide up to a boundary term and the equations of motion for both theories are of second-order, which suggests that they are equally hard to quantise in either formulation.

\smallskip

While there is the duality \eqref{tel-GR} between the teleparallel action and the GR action in the second-order formalism, to our knowledge, there is no similar derivation of teleparallel gravity from a first-order action. Indeed, having a zero-curvature connection would solve the equation of motion (when there is no massive degrees of freedom), but plugging this back into the action \eqref{actionEC} yields a totally vanishing action, which is undesirable. Some ways to avoid a vanishing action is to either supplement \eqref{actionEC} with a constraint implementing the zero-curvature constraint \cite{Blagojevic:2000qs}, or even to add quadratic contributions in torsion and curvature \cite{Baekler:2011jt}. 

\smallskip

Recent developments in loop quantum gravity (LQG)  indicate that there ought to be also a first-order formulation of teleparallel gravity and a more symmetric treatment of the GR and teleparallel formulations at the quantum level. \footnote{See also \cite{Mielke:1992te} which used the Ashtekar variables in the teleparallel context. } The standard LQG framework is based on imposing the Gauss constraint first, which amounts---at least in three spacetime dimensions---to deal with a torsionless connection. At the quantum level, this leads to the so-called Ashtekar--Lewandowski vacuum. From this perspective, LQG can be seen as quantization of the metric GR formalism  \cite{Thiemann:2007pyv}.

\smallskip

Dittrich and Geiller suggested that there should be another interesting realisation of quantum geometry to start with, not based on the imposition of the Gauss constraint first, but instead on a zero curvature constraint. At the quantum level, this leads to the so-called BF vacuum \cite{Dittrich:2014wda,Dittrich:2014wpa}, which resembles the teleparallel formulation since it is based on a global notion of flat parallel transport. The idea that the BF vacuum provides a realisation of  teleparallel gravity at the quantum level is strengthened by two recent developments: it was shown in \cite{Dupuis:2017otn, Shoshany:2019ymo} that there are two natural ways to discretize the Einstein--Cartan gravity action. The first way is to start with the Gauss constraint (this is the \emph{\q{loop gravity}} framework), the other way  implements a zero curvature constraint first (this is the \emph{\q{dual loop gravity}} framework). The dual framework provides  a semi-classical realisation of the idea suggested by Dittrich and Geiller \cite{Dittrich:2014wda}. In each case, the starting point is the classical BF type of gravitational action. However, it was not shown explicitly that the dual loop gravity framework is related to teleparallelism. Two further developments provide additional evidence in favour of such a relation: in \cite{Tiede:2017fch}, it was argued that t'Hooft's discrete approach to three-dimensional gravity can be seen as a discretization of the teleparallel formulation, and in \cite{clement}, which builds upon the results of Dittrich and Geiller, a quantization of a dual loop gravity model was developed, which led to the Dijkgraaf--Witten model.

%

\smallskip

Many arguments point, therefore, to the fact that the teleparallel formulation should also be present in the Einstein--Cartan formulation. In this  note, we want to  illustrate how this could be achieved at least in three spacetime dimensions. In section \ref{firstorder}, we show that the Einstein--Cartan first-order formulation of the standard GR theory, is also a first-order formulation of the teleparallel theory up to a boundary term. This is done first in the three-dimensional Euclidean case where the Einstein-Cartan action is simply the $\SU(2)$ BF action. We then generalize our derivation to arbitrary dimensions. The main idea of the derivation is to decompose the off-shell spin connection into a fiducial reference connection plus a difference tensor $\ou{\Delta}{I}{Ja}$. Depending on whether the reference connection is flat or given by the Levi-Civita connection, the Einstein--Cartan action is equal to either the Palatini action of GR or the teleparallel action plus a boundary term.

We will then discuss  the different discretizations performed  in \cite{Dupuis:2017otn, Shoshany:2019ymo} (for three-dimensional gravity) in light of the observation that both the GR and teleparallel frameworks can be derived from the same first-order action (up to a boundary term). In the Hamiltonian picture, each of these frameworks can be naturally associated to a choice of polarization. The physical equivalence of the different polarizations is the mathematical manifestation of the physical equivalence between the GR and teleparallel frameworks.  We will argue however that different choices of polarization at the continuum level lead to different discretized theories. More explicitly, the choice of polarization in the continuum and the discretization procedure used in \cite{Dupuis:2017otn, Shoshany:2019ymo} do not commute. Finally, we will discuss how the dual loop gravity picture can be seen as a discretized version of the teleparallel formulation.

%
%


\section{First order action for teleparallel gravity }\label{firstorder}
\noindent We detail below how the Einstein--Cartan action \eqref{actionEC} can be seen as the first-order formulation of the teleparallel action. We first focus  on the three-dimensional Euclidean case as a warm-up. Three-dimensional Euclidean gravity is very well understood in the loop quantum gravity framework.
We then study the general $D$-dimensional Lorentzian case.

\smallskip

The key idea is based on the observation that the space of $SO(1,D-1)$ connections is an affine space: any two spin connections are separated by a difference tensor $\ou{\Delta}{I}{Ja}$. As a consequence, any connection $\ou{A}{I}{Ja}$ can be parametrised in terms of a reference connection $\ou{\omega}{I}{Ja}$, which provides an arbitrary origin in the affine space of connections, plus a difference tensor $\ou{\Delta}{I}{Ja}$ that encodes the dynamical degrees of freedom. In the absence of symmetries, there are only two natural choices (modulo internal $SO(1,D-1)$ gauge transformations) for such a  metric-compatible reference connection, namely the Weitzenb\"ock connection, $\oo$, and the Levi-Civita connection $\oLC$. They respectively have   no curvature or no torsion.

Solving the equations of motion for the difference tensor $\Delta$ will allow us to re-express the Einstein--Cartan action \eqref{actionEC} as the teleparallel action provided the reference connection is the Weitzenb\"ock connection, while the other choice gives the standard GR case. 

\subsection{The BF action in three dimensions} \label{3dcase1}
\noindent The starting point is the Einstein-Cartan action \eqref{actionEC} for three-dimensional Euclidean gravity.  
\be \label{3dBFaction}
S_{\mathrm{EC}}[e,A]=-\frac{1}{8\pi G}\int_{\mathcal{M}} \big\langle e \wedge F[A] \big\rangle , \quad F[A]=\rd A + \demi [A, A],
\ee
 where $\la X,Y\ra=X_IY^I$ is the Killing form for $\su(2)$, and both the frame field $e$ and the connection $A$ are with value in $\su(2)$. 
Taking into account the parametrisation $\ou{A}{I}{a}=\ou{\omega}{I}{a}+\ou{{\Delta}}{I}{a}$ of the connection into an arbitrary fiducial reference connection $\ou{\omega}{I}{a}$ and a displacement vector $\ou{\Delta}{I}{a}$, the $\SU(2)$ field strength becomes
\begin{align}\nonumber
	F[A]&=\di A+\frac{1}{2}\big[A,A\big]=\di\omega+\frac{1}{2}\big[\omega,\omega\big]+\di\Delta+[\omega,\Delta]+\frac{1}{2}\big[\Delta,\Delta\big]\\
	&=F[\omega]+\prefix{\omega\!}{D}\Delta+\frac{1}{2}\big[\Delta,\Delta\big].\label{curvtr1}\end{align}
where $\prefix{\omega\!}{D}=\di+[\omega,\cdot]$ is the exterior covariant derivative with respect to the reference connection.
At the level of the action, we thus have,
\bes
\label{split action in diff. form 1}
S_{\mathrm{EC}}[e,A] = S_{\mathrm{EC}}[e,\Delta;\omega]
=-\frac{1}{8\pi G}\int_{\mathcal{M}}\Big\langle e\wedge F[\omega]-\di(e\wedge\Delta)+\prefix{\omega}{T}\wedge \Delta+\frac{1}{2}e\wedge[\Delta\wedge\Delta]\Big\rangle
\ees
where $\prefix{\omega}{T}=\di e+[\omega,e]$ is the torsion of the connection $\omega$. The second term is a total exterior derivative, using Stokes's theorem it turns into a surface integral. 

Let us then consider the case where $\omega=\oo$, which by definition of the Weitzenb\"ock connection $\oo$  is such that $F[\oo]=0$.
 Hence the first term in the action \eqref{split action in diff. form 1} vanishes. Up to a boundary term, we then have 
\bes
\label{contorsion action}
S_{\mathrm{EC}}[e,\Delta;\prefix{\bullet}{\omega}] 
&=& - \frac{1}{8\pi G}\int_{\mathcal{M}}\Big\langle \bT \wedge\Delta  + \frac{1}{2}e\wedge[\Delta\wedge\Delta]\Big\rangle, 
\ees 
where we denoted $\bT=\bD e=\rd e+[\oo,e]$. 
Variations in terms of $e$ and $\Delta$ respectively give, 
\bes\label{einstein2}
&& \bD\Delta +\frac{1}{2} [\Delta\wedge\Delta] =0\\
\label{contorsion equation}
&& \bT  + [ e\wedge\Delta]= 0.
\ees
Provided the frame field is invertible, we can solve the last equation of motion, and  express the difference tensor $\Delta$ in terms of the frame field and the torsion tensor $\prefix{\bullet}{T}{}^I{}_{ab}$ 
associated to the Weitzenb\"ock connection. 
\be
\ou{\Delta}{I}{a} =-\demi\;\epsilon^{I}{}_{JK}\left(e^{bJ}\;\bT{}^{K}{}_{ab}-\demi\;e^{c J }\;e^{b K}\;\bT{}_{acb}\right),
\ee 
where $\pou{\bullet}{T}{a}{bc}=\uo{e}{I}{a}\pou{\bullet}{T}{I}{bc}$.
We can now plug this expression back in the action \eqref{contorsion action}. After some algebra, we recover the teleparallel action \cite{tele}.
\begin{align}\nonumber 
S_{\mathrm{EC}}&[e,A]-\frac{1}{8\pi G}\int_{\partial\mathcal{M}}\big\langle e\wedge \Delta\big\rangle\\\label{teleparal}
&\approx-\frac{1}{16\pi G}\int d^3v_e\left(\frac{1}{4}\;\bT^{a}{}_{bc}\;\bT_{a}{}^{bc} - \demi\;\bT^{c}{}_{ab}\;\bT^{ab}{}_{c} -\;\bT^{c}{}_{bc}\;\bT^{ab}{}_{a}\right)=: S_{\mathrm{TP}}^{\mathrm{eucl.}}[e;\oo],
\end{align}
where $\approx$ means that we went on-shell in terms of the equation of motion for $\Delta$, and $d^3v_e=\frac{1}{6}\epsilon_{IJK}e^I\wedge e^J\wedge e^K$ is the three-volume element.

\smallskip

The Einstein--Cartan action is therefore a first-order formulation of teleparallel gravity. As we have just shown the standard teleparallel action is recovered by choosing as reference connection $\omega$ the Weitzenb\"ock connection $\oo$ and by plugging back the equations of motion coming from the variations with respect to $\Delta$ into the Einstein--Cartan action. The equality between the two actions \eqref{teleparal} is indeed valid up to a boundary term and on-shell. 

 \medskip

Choosing a different reference connection, and repeating the same steps as before to eliminate $\Delta$, we obtain the Palatini action. Let us briefly sketch the different steps. First of all, we take the reference connection $\omega$ to be the Levi-Civita connection $\oLC$, which is such that $\ciT=\ciD e=0$, where $\ciD = \di +[\oLC,\cdot]$ is the covariant exterior derivative with respect to the Levi-Civita connection.  The action \eqref{3dBFaction} then becomes 
\bes
\label{split action in diff. form 2}
S_{\mathrm{EC}}[e,A] = S_{\mathrm{EC}}[e, \oLC+\Delta]
&=&-\frac{1}{8\pi G} \int_{\mathcal{M}}\Big\langle e\wedge F[\oLC] + e \wedge \prefix{\circ}D\Delta+ \frac{1}{2} e\wedge[\Delta\wedge\Delta]\Big\rangle .
\ees
Variations along $\Delta$ give $[e,\Delta]=0$. Assuming again that $e$ is invertible,  the solution of this equation is given by $\Delta=0$. Plugging back this solution in \eqref{split action in diff. form 2} yields the Palatini action for three-dimensional gravity in the second-order formalism, 
\bes\label{pal3d}
S_{\mathrm{EC}}[e,A] \approx S_{\mathrm{EC}}[e,\oLC]
&=&-\frac{1}{8\pi G}  \int\big\langle e\wedge F[\oLC]\big\rangle \equiv S_{\text{Palatini}}[e].
\ees 

\subsection{Teleparallel gravity in $D$ dimensions from $D$-dimensional Einstein--Cartan action}
\noindent The same construction holds in $D$ Lorentzian spacetime dimensions. Consider the Einstein--Cartan action
\begin{equation}
S_{\mathrm{EC}}[A,e]=\frac{1}{16\pi G}\int_{\mathcal{M}}B_{IJ}[e]\wedge F^{IJ}[A], \label{actndef1}
\end{equation}
where $\ou{F}{I}{J}$ is the curvature two-form
\begin{equation}
\ou{F}{I}{J}=\di\ou{A}{I}{J}+\ou{A}{I}{M}\wedge \ou{A}{M}{J},
\end{equation}
and $B_{IJ}$ is the bivector-valued $(D-2)$-form
\begin{equation}
B_{IJ}=\frac{1}{(D-2)!}\epsilon_{IJK_1\dots K_{D-2}}e^{K_1}\wedge\dots\wedge e^{K_{D-2}}.
\end{equation}
To write the action in a more familiar form, we decompose the curvature two-form into its components with respect to the $D$-bein, namely
\begin{equation}
\ou{F}{I}{J}=\frac{1}{2}\ou{F}{I}{JKL}[A,e]\,e^K\wedge e^L,
\end{equation}
which is possible as long as the $D$-bein is invertible. A short calculation gives,
\begin{equation}
S_{\mathrm{EC}}[A,e]=\frac{1}{16\pi G}\int_{\mathcal{M}} d^Dv_e\,\ou{F}{IJ}{IJ}[A,e],\label{EHactn2}
\end{equation}
where we introduced the $D$-dimensional volume element,
\begin{equation}
d^Dv_e=\frac{1}{D!}\epsilon_{I_1\dots I_D}e^{I_1}\wedge\dots \wedge e^{I_D}.
\end{equation}
Let us now explain how to recover the GR and teleparallel formulations. Consider first an arbitrary origin $\ou{\omega}{I}{J}$ in the affine space of connections and parametrize any connection in terms of $\ou{\omega}{I}{J}$ and a displacement vector $\ou{\Delta}{I}{J}$, which is an $\mathfrak{so}(1,D-1)$-valued one-form. Thus,
\begin{equation}
\ou{A}{I}{J}=\ou{\omega}{I}{J}+\ou{\Delta}{I}{J}.
\end{equation}
Let now $\prefix{\omega}{D}$ denote the exterior covariant derivative with respect to  $\,\ou{\omega}{I}{J}$. The curvature two-form satisfies
\begin{equation}
\ou{F}{I}{J}[A]=\ou{F}{I}{J}[{\omega}]+\prefix{\omega}{D}\ou{\Delta}{I}{J}+\ou{\Delta}{I}{L}\wedge\ou{\Delta}{L}{J}.
\end{equation}
If $\ou{\omega}{I}{Ja}$ is the torsionless Levi-Civita spin connection $\pou{\circ}{\omega}{I}{Ja}$, the corresponding curvature two-form is nothing but the Riemann curvature tensor. In components,
\begin{equation}
\ou{F}{I}{Jab}[\prefix{\circ}{\omega}]=\ou{e}{I}{c}\uo{e}{J}{d}\ou{R}{c}{dab}[g].
\end{equation}
In this case, the action \eref{EHactn2} reduces, therefore, to the usual metrical Einstein--Hilbert action, provided $\Delta=0$.

\medskip

If we are interested in teleparallel gravity, the relevant reference connection is the Weitzenb\"ock connection $\prefix{\bullet} \omega$, which has vanishing curvature. Performing a partial integration, we are then left with the following expression for the action,
\begin{align}\nonumber
S_{\mathrm{EC}}[A,e]+(-1)^{D-1}\int_{\partial\mathcal{M}}&B_{IJ}\wedge\Delta^{IJ}=\\
\nonumber=\frac{1}{16\pi G}\int_{\mathcal{M}}\bigg[&\frac{1}{(D-3)!}\epsilon_{L_1\dots L_{D-3}IJK}e^{L_1}\wedge\dots\wedge e^{L_{D-3}}\wedge\prefix{\bullet}{T}^I\wedge\Delta^{JK}+\\
&+\frac{1}{(D-2)!}\epsilon_{L_1\dots L_{D-2}IJ}e^{L_1}\wedge\dots\wedge e^{L_{D-2}}\wedge\ou{\Delta}{I}{L}\wedge\Delta^{LJ}\bigg].\label{actndef2}
\end{align}
where we introduced the Weitzenb\"ock torsion, 
\begin{equation}
\po{\bullet}{T}{I}=\prefix{\bullet}{D} e^I.
\end{equation}
The algebraic structure of the action \eref{actndef2} can be considerably simplified by noting that
\begin{equation}
\epsilon_{L_1\dots L_{d-n}I_1\dots I_n}\epsilon^{L_1\dots L_{d-n} J_1\dots J_n}=-n! (d-n)! \delta^{[J_1}_{I_1}\dots \delta^{J_n]}_{I_n}.
\end{equation}
Consider then the components of the Weitzenb\"ock torsion with respect to the $D$-bein,
\begin{equation}
\prefix{\bullet}{T}^I=\prefix{\bullet}{D}e^I=\frac{1}{2}\ou{\prefix{\bullet}{T}}{I}{LM}e^L\wedge e^M.
\end{equation}
This allows us to write the action  \eref{actndef2} in the following compact form
\begin{align}
\nonumber S_{\mathrm{EC}}[A,e]+&\frac{(-1)^{D-1}}{16\pi G}\int_{\partial\mathcal{M}}B_{IJ}\wedge\Delta^{IJ}=\\
\nonumber=&\frac{1}{16\pi G}\int_{\mathcal{M}}d^Dv_e\bigg[3\pou{\bullet}{T}{I}{MN}\ou{\Delta}{JK}{R}\delta^{[M}_I\delta^{N\phantom{]}}_J\delta^{R]}_K+2\delta^{[M}_I\delta^{N]}_J\ou{\Delta}{I}{LM}\ou{\Delta}{LJ}{N}\bigg]\\
=&\frac{1}{16\pi G}\int_{\mathcal{M}}d^Dv_e\bigg[2\pou{\bullet}{T}{M}{MJ}\ou{\Delta}{JN}{N}+\prefix{\bullet}{T}_{IJK}\Delta^{JKI}- \ou{\Delta}{I}{LI}\ou{\Delta}{JL}{J}-\ou{\Delta}{N}{[LM]}\ou{\Delta}{LM}{N}\bigg].\label{KT}
\end{align}
where we decomposed the difference tensor one-form $\ou{\Delta}{I}{Ja}$ into its components $\ou{\Delta}{I}{Ja}=\ou{\Delta}{I}{JM}\ou{e}{M}{a}$ with respect to the $D$-bein $e^I$. 

To express this action in terms of the torsion two-form alone, we have to impose strongly the torsionless condition at the level of the action. In other words, part of the equations of motion are plugged back into the action to eliminate $\Delta$ as an independent variable. Consider first the variation of the action with respect to the difference tensor one-form $\ou{\Delta}{I}{Ja}$, which yields the  condition,
\begin{equation}\label{trsnless1}
\prefix{\bullet}{T}^I+\ou{\Delta}{I}{J}\wedge e^J=0.
\end{equation}
In terms of its components, the  condition \eqref{trsnless1} is now solved by
\begin{equation}
\Delta_{IJK}=-\Delta_{JIK}=\frac{1}{2}\big({\prefix{\bullet}{T}}_{IJK}+{\prefix{\bullet}{T}}_{JKI}-{\prefix{\bullet}{T}}_{KIJ}\big).\label{KT1}
\end{equation}
This in turn implies
\begin{subequations}
\begin{align}
\ou{\Delta}{N}{[LM]}&=\frac{1}{2}\pou{\bullet}{T}{N}{LM},\label{KT2}\\
\ou{\Delta}{JN}{N}&=\puo{\bullet}{T}{N}{NJ}.\label{KT3}
\end{align}
\end{subequations}
If we now insert (\ref{KT1}, \ref{KT2}, \ref{KT3}) back into \eref{KT}, we get the usual teleparallel action which is now quadratic in the components of the torsion two-form,
\begin{align}
\nonumber S[A,e]+&\frac{(-1)^{D-1}}{16\pi G}\int_{\partial\mathcal{M}}B_{IJ}\wedge{\Delta}^{IJ}\approx\\
&\approx\frac{1}{16\pi G}\int_{\mathcal{M}}d^Dv_e\bigg[\pou{\bullet}{T}{M}{MJ}\puo{\bullet}{T}{N}{NJ}+\frac{1}{2}\prefix{\bullet}{T}_{NLM}\prefix{\bullet}{T}^{LMN}-\frac{1}{4}\prefix{\bullet}{T}_{NLM}\prefix{\bullet}{T}^{NLM}\bigg],
\end{align}
where $\approx$ denotes terms that vanish provided the torsionless condition \eref{trsnless1} is satisfied. 

As in the three-dimensional Euclidean case, we have proved in the Lorenztian $D$-dimensional case that the Einstein--Cartan action, a well-known first-order formulation of the standard GR formulation (Palatini action), is also a first-order formulation of the teleparallel action  up to a boundary term.

\section{Relating the dual loop picture to the teleparallel formulation in three dimensions} \label{symplectic}
\noindent We now focus on the three-dimensional Euclidean case, and restrict ourselves to a trivial topology $\mathcal{M}\sim \mathbb{R}\times \Sigma$, with  the spatial manifold $\Sigma$ having no boundary for simplicity. As in section \ref{3dcase1}, the fundamental configuration variables, namely the triad $e$ and the connection $A$, are one-forms with value in $\su(2)$. 

We will show that  starting from the Einstein-Cartan action there are two natural symplectic potentials that appear, related by an integration by parts. They amount to different choices of polarization. Following our previous result, namely that the Einstein--Cartan action can be seen as the first-order action of both  GR and teleparallel gravity, we will argue that the different choices of polarization are naturally related to  either the GR or teleparallel frameworks. 

We will then recall how the discretization procedure described in \cite{Dupuis:2017otn, Shoshany:2019ymo} gives rise to different discrete theories. Each discrete theory can be then naturally identified with the different choices of polarization in the continuum. Hence we will argue that the dual loop gravity discrete theory can naturally be seen as a discretization of the teleparallel framework.

\subsection{Pre-symplectic forms in the continuum}\noindent 
Standard calculations for the Einstein-Cartan action
\be \label{actionBF3}
S_{\mathrm{EC}}[e,A]= -\frac{1}{8\pi G}\int_{\mathcal{M}} \big\langle e\wedge F[A]\big\rangle,
\ee
 lead to the pre-symplectic potential 
\be\label{tbf}
\Theta_{\mathrm{EC}}= -\frac{1}{8\pi G}\int_\Sigma \uo{\tilde{E}}{I}{a}\delta\ou{A}{I}{a},
\ee
where $\delta$ is the differential in field space and $\uo{\tilde{E}}{I}{a}$ denotes the densitized triad\footnote{In the following, indices $a,b,c,\dots$ are two-dimensional abstract tensor indices and $\tilde{\varepsilon}^{ba}$ is the Levi-Civita skew-symmetric tensor density on the spatial slice.}
\begin{equation}
\uo{\tilde{E}}{I}{a}=\tilde{\varepsilon}^{ab}e_{Ib}.
\end{equation}
On the other hand, we now also have on field space
\begin{equation}
\delta\ou{A}{I}{a}=\delta\big[\ou{\omega}{I}{a}+\ou{\Delta}{I}{a}\big]=\delta \ou{\Delta}{I}{a},
\end{equation}
since ${\omega}$ is a reference connection, which is kept fixed on field space.
\be\label{tbfbis}
\Theta_{\mathrm{EC}}= \frac{1}{8\pi G}\int_\Sigma \uo{\tilde{E}}{I}{a}\delta\ou{\Delta}{I}{a},
\ee

Let us now choose the Weitzenb\"ock connection $\prefix{\bullet}\omega$ as the reference connection. Then,  the action \eqref{actionBF3} becomes, up to a boundary term, \eqref{contorsion action} 
\be \label{contorsion again}
S_{\mathrm{EC}}[e,\prefix{\bullet}{\omega}]= - \frac{1}{8\pi G}\int_{\mathcal{M}}\Big\langle \bT \wedge\Delta  + \frac{1}{2}e\wedge[\Delta\wedge\Delta]\Big\rangle.
\ee
We refer to section \ref{3dcase1} for more details. Direct calculations lead this time to the symplectic potential 
\be\label{ptp}
\Theta_{\mathrm{TP}}=- \frac{1}{8\pi G}\int_\Sigma \big\langle \delta e\wedge \Delta\big\rangle=-\frac{1}{8\pi G}\int_\Sigma\tilde{\varepsilon}^{ab}\delta e_{Ia}\,\ou{\Delta}{I}{b}=-\frac{1}{8\pi G}\int_\Sigma \uo{\tilde{\Delta}}{I}{a}\delta\ou{e}{I}{a},
\ee
where we introduced the densitized difference tensor
\begin{equation}
\uo{\tilde{\Delta}}{I}{a}={\tilde{\varepsilon}}^{ab}\Delta_{Ib}.
\end{equation}
We refer to this symplectic potential as the symplectic potential for the teleparallel picture since \eqref{contorsion again} is the teleparallel action \eqref{teleparal} on-shell.
\medskip

The actions  \eqref{actionBF3} and \eqref{contorsion again}  are related by an integration by part. The relevant connection variables for the symplectic form are actually given in terms of the difference tensor. Since we are dealing with densitized fields, the canonical map between the two polarizations also implements a (Poincar\'e) dualization via the Levi-Civita tensor density $\tilde{\varepsilon}^{ab}$,  
\be
(\tilde{\varepsilon}^{ab}e_{Ib}, \ou{\Delta}{I}{a})\rightarrow (\tilde{\varepsilon}^{ab}\Delta_{Ib}, \ou{e}{I}{a}).
\ee
These two sets of variables represent two polarisations on the classical phase space, corresponding to either the GR formulation or the teleparallel formulation in the continuum. The physics must be independent of the choice of polarization. 
This is another way to say that  we can equivalently work with the GR or teleparallel formulations.  

Hence from an abstract perspective, the choice of polarization does not matter at the continuum level. 
At the discrete level, however, things will be more subtle. Indeed, the discretization procedure is sensitive to the dualization induced by  the  Levi-Civita tensor density $\tilde{\varepsilon}^{ba}$. Let us explain this point below.

\subsection{Symplectic forms in the discrete picture}
\noindent We recall the construction of \cite{Dupuis:2017otn}, neglecting  the possible existence of curvature and torsion defects at the vertices of the triangulation. For further details about such issues, see  
\cite{Freidel:2018pbr}.
\medskip 

The phase space underlying the spin network quantum states can be obtained through a discretization procedure, which relies on two steps. The first step is to discretize the spatial manifold using a triangulation. The second step is to truncate  the degrees of freedom by assuming that on the faces $c^*$ of the triangulation  we have the constraints satisfied, meaning that there is neither torsion  nor curvature inside $c^*$. 
The solutions of such zero torsion and zero curvature constraints are respectively given by 
\be \label{sol00}
e(x)=g_c^{-1}\rd y_cg_c,\quad A= g_c^{-1}\rd g_c, 
\ee
with $x$ any point of a given face $c^*$ of the triangulation, $g_c(x)$ the holonomy joining the reference point $c$ to $x$ in $c^*$, and $y_c$ a Lie algebra element. 

We intend to discretize the pre-symplectic potential $\Theta_{\text{EC}}$ \eqref{tbfbis} rather than $\Theta_{\text{TP}}$ \eqref{ptp}, as the latter cannot be written in terms of boundary data only. Nevertheless, we will still be able to have the discrete analogue of the potential  $\Theta_{\text{TP}}$ \eqref{ptp} precisely because the discretized version  of $\Theta_{\text{EC}}$ \eqref{tbfbis} will be an exact two-form, essentially allowing for an integration by parts that relates the discretization of $\Theta_{\text{EC}}$ to a discrete version of $\Theta_{\text{TP}}$.

\smallskip

Starting from $\Theta_{\text{EC}}$ \eqref{tbfbis}, within a face $c^*$ of the triangulation, we replace the frame field and the connection by their respective discrete expression given in \eqref{sol00}
\be
\Theta_{\text{EC}}=\frac{1}{8\pi G}\int_{c^*} \big\langle e\wedge \delta A\big\rangle = \frac{1}{8\pi G}\int_{c^*} \left\langle\rd y_c \wedge \rd (\delta g_cg_c^{-1})\right\rangle. 
\ee
As the integrand is an exact two-form,  this integral can be evaluated on the boundary of $c^*$ and  there are two possible choices to do so.
 \be
  \int_{c^*} \big\langle\rd y_c \wedge \rd (\delta g_cg_c^{-1})\big\rangle= - \int_{\partial c^*}\big\langle \rd y_c \,  (\delta g_cg_c^{-1})\big\rangle= \int_{\partial c^*}\big\langle y_c \,  \rd (\delta g_cg_c^{-1})\big\rangle.
\ee
Such discretization can be performed for any face, in particular for the face $c'^*$ which shares an edge $\ell$ as boundary with $c^*$. Furthermore the fields $g_{c'}(x)$ and $y_{c'}(x)$ being evaluated on $\ell$ can be related to the fields $g_{c}(x)$ and $y_{c}(x)$ evaluated at the same point on $\ell$ . 
\be
g_{c'}=h_{c'c}g_c, \quad y_{c'}= h_{c'c}(y_c+x_{cc'})h_{c'c}^{-1}.
\ee
These are the continuity conditions at $\ell$, the common edge of the faces $c^*$ and $c'^*$. Implementing these relations for each contributions $c^*$, $c'^*$ for the edge $\ell=[vv']$, which is dual to the spin network link $[cc']=\ell^*$, we get the two different potentials, for each edge $\ell$.
\bes
&&\Theta^\ell_{\text{LG}}=- \frac{1}{8\pi G}\Big\langle\left(\int_\ell \rd y_c \right) \delta h_{\ell^*}h_{\ell^*}^{-1}\Big\rangle= -\frac{1}{8\pi G}\Big\langle X_{\ell}\, \delta h_{\ell^*}h_{\ell^*}^{-1}\Big\rangle, \label{sy1}\\
&&\Theta^\ell_{\text{LG}^*}= +\frac{1}{8\pi G}\Big\langle(g_{vc}x_{\ell^*}g_{cv})\delta g_{\ell}g_{\ell}^{-1}\Big\rangle= +\frac{1}{8\pi G}\Big\langle X_{\ell^*} \, \delta g_{\ell}g_{\ell}^{-1}\Big\rangle, \label{sy2}
\ees
where we used the notations 
\bes
X_\ell \equiv \int_\ell \rd y_c, &\quad&   h_{\ell^*}\equiv g_{cv}g_{vc'} \\
 X_{\ell^*}\equiv g_{vc}x_{\ell^*}g_{cv}, &\quad&  g_\ell\equiv g_{vc}g_{cv'}. 
\ees
$\Theta^\ell_{\text{LG}}$ refers to the loop gravity potential,  whereas $\Theta^\ell_{\text{LG}^*}$ refers to the dual loop gravity potential. 

By construction, in \eqref{sy1}, the fluxes $X_{\ell}$ satisfy the Gauss constraint when summing over the edges of a given triangle.
\be
\sum_{\ell\in\partial c^*} X_\ell =0.
\ee
This is the discretized version of the torsionless condition (a two-form) on a triangle. 
The data $(X_\ell, h_{\ell^*}, \Theta^\ell_{\text{LG}})$ provides the classical phase space for the standard spin networks: we have holonomies decorating the dual of the triangulation, i.e.\ the spin network graph. This is often coined the \textit{loop gravity framework.}

\medskip 
On the other hand we also have the dual picture  \eqref{sy2}  where the holonomies $g_\ell$ around the triangles satisfy the flatness constraint,
\be
\prod_{\ell\in\partial c^*}g_{\ell}=1.
\ee
This is the discretized version of the constraint that imposes that the curvature is flat. The data $(g_{\ell}, X_{\ell^*}, \Theta^\ell_{\text{LG}^*})$ provides the classical phase space for the \emph{\q{dual}} spin networks: we have fluxes decorating the dual of the triangulation, i.e.\ the spin network graph. This is  naturally coined \textit{dual loop gravity}. Such a discrete theory was shown to be related to t'Hooft approach to three-dimensional gravity  \cite{Tiede:2017fch} and the Dijkgraaf-Witten model \cite{clement}. 

\medskip

The parallel with the previous section should now be clear.  The configuration variables $\ou{\Delta}{I}{a}$, $ \ou{e}{I}{a}$,  are discretized along the link $\ell^*$, whereas the momentum variables $\uo{\tilde{E}}{I}{a}$, $\uo{\tilde{\Delta}}{I}{a}$,  are discretized along the edge $\ell$.\vspace{-1em}
\begin{center}
\begin{tabular}{ccc}
&\hspace{1cm} & \\
\begin{tabular}{rcl}
\emph{\q{GR polarization}} &$\dr$& loop gravity\\\hline
 $\uo{\tilde{E}}{I}{a} $ &$\dr$ &$X_\ell$\\
$\ou{{\Delta}}{I}{a}$&$\dr$ &$h_{\ell^*}$\\
 $\Theta_{\text{EC}}$&$\dr$ &$\Theta^\ell_{\text{LG}}$
\end{tabular} & &
\quad\begin{tabular}{rcl}
\emph{\q{Teleparallel polarization}} &$\dr$& dual loop gravity\\ \hline
$\uo{\tilde{\Delta}}{I}{a}$&$\dr$ & $g_{\ell}$ \\
$\ou{e}{I}{a}$&$\dr$ & $X_{\ell^*}$\\
$\Theta_{\text{TP}}$&$\dr$ & $\Theta^\ell_{\text{LG}^*}$
\end{tabular}
\end{tabular}
\end{center}\vspace{0.5em}
Dual loop gravity can be interpreted as the discretization of the teleparallel framework, just like loop gravity can be seen as a discretization of GR.

The momentum variables are discretized on structures dual to the ones which the configuration variables are associated to. Hence to different polarizations in the continuum are associated \textit{different} discretizations. Change of polarization at the continuum level and discretization do not commute.
 
Yet, the physics cannot depend on the choice of polarization, and different discretisations should not lead to different physics either.   Ignoring issues with the continuum limit \cite{Dittrich:2014ala}, which can be set aside in three-dimensional gravity, because the theory is topological, we expect, therefore, that the two discretizations must be related by a duality map, encoding their equivalence. Such a duality was conjectured in \cite{clement} and is probably related to the one found in the context of the Kitaev model \cite{Buerschaper2010}. We will  leave this for further investigations.
   
\section*{Discussion}
\noindent To the best of our knowledge, until now there was no proposal for a first-order formulation for teleparallel gravity.  In the second-order formalism, the teleparallel action is obtained from the usual GR action by an integration by part. It is not so surprising to see, therefore, that the first-order action for teleparallel gravity is obtained from an integration by parts from the first-order Einstein--Cartan action for gravity.  The key idea is to parametrise the spin connection degrees of freedom in terms of a fiducial reference connection, which is undynamical, and a difference tensor\footnote{If the reference connection were the Levi-Civita connection, the difference tensor would be nothing but the contortion tensor.} $\Delta$ that encodes the dynamical degrees of freedom. Then, this parametrisation allows us to show that the Einstein-Cartan action provides the first-order formulation of \emph{both} metric general relativity as well as the first-order formulation of  teleparallel gravity. 
 
This result allows us to justify the statement made in \cite{Dupuis:2017otn}. Namely that dual loop gravity is related to the teleparallel picture and that furthermore the  loop gravity and dual loop gravity representations are related by a change of polarization. 
 These two polarizations are equivalent in the continuum but lead to two different discrete theories.  We expect that the equivalence between these different polarisations should be realised as an equivalence between different dual lattice regularisations  (implementing the  Poincar\'e duality found in the continuum). This is currently being investigated.   

Another interesting question is to understand how the cosmological constant modifies the construction described in this article. From the three-dimensional quantum gravity side, it is well known that a quantum group structure emerges. This can be traced back to the fact that we discretise the theory using homogeneously curved geometries.  On the other hand Dittrich and Geiller \cite{Dittrich:2016typ} discussed how the dual BF vacuum construction is also deformed using quantum group structures. This suggests that there must be a teleparallel formulation of gravity that is discretized along some teleparallel analogue of homogeneously curved geometries.    We leave this intriguing question for further investigations.

 \subsection*{Acknowledgements}
\noindent  We thank Laurent Freidel for discussions. This research was supported in part by Perimeter Institute for Theoretical
Physics. Research at Perimeter Institute is supported by the Government
of Canada through the Department of Innovation, Science and Economic
Development Canada  and by the Province of Ontario through the Ministry
of Research, Innovation and Science. A.\,O.\ is supported by the NSERC Discovery grants held by M.\,D.\ and F.\,G. 
 
 \bibliographystyle{Utphys}
\phantomsection\addcontentsline{toc}{section}{\refname}\bibliography{tele}

 \end{document}